\documentclass[twocolumn,11pt]{article}
\usepackage{times}
\usepackage{graphicx}
\begin{document}
\global\def\refname{{\normalsize \it References:}}
\baselineskip 12.5pt
%
%
% TITLE, AUTHOR, ABSTRACT, KEYWORDS
%
\title{\LARGE \bf The effect of limiting resources in aging populations}

\date{}

\author{\hspace*{-10pt}
\begin{minipage}[t]{2.7in} \normalsize \baselineskip 12.5pt
\centerline{Chrysline Margus Pi\~nol}
\centerline{University of the Philippines}
\centerline{National Institute of Physics}
\centerline{Diliman 1101 Quezon City, Philippines}
\centerline{and}
\centerline{University of the Philippines}
\centerline{Institute of Mathematical Sciences and Physics}
\centerline{Los Ba\~nos 4031 Laguna, Philippines}
\centerline{cpinol@nip.upd.edu.ph}
\end{minipage} \kern 0in
\begin{minipage}[t]{2.7in} \normalsize \baselineskip 12.5pt
\centerline{Ronald Banzon}
\centerline{University of the Philippines}
\centerline{National Institute of Physics}
\centerline{Diliman 1101 Quezon City, Philippines}
\centerline{rbanzon@nip.upd.edu.ph}
\end{minipage}
%
% If you are three authors then you can use three mini--pages
% instead of two. Their horizontal size must be less than 2.7in
% indicated above. It can be e.g. 2.3in. However, you must pay
% attention that you do not exceed the total width of the text.
%
\\ \\ \hspace*{-10pt}
\begin{minipage}[b]{6.9in} \normalsize
\baselineskip 12.5pt {\it Abstract:}
% The text of the abstract follows.
The concept of a carrying capacity is essential in most models to prevent unlimited growth. Despite the large amount of deaths it introduces, the actual influence of the Verhulst term in simulations is often times not accounted for. Generally, it is treated merely as a scaling parameter that functions to keep simulated populations within computer limits. Here, we compare two different implementations of the concept in the Penna model - Vehulst applied to all individuals (VA) and to newborns only (VB). We observe variations in certain model features when random Verhulst deaths are restricted to a single age group.
\\ [4mm] {\it Key--Words:}
% The key-words follow.
Population dynamics, Carrying capacity, Aging, Penna model
\end{minipage}
\vspace{-10pt}}

\maketitle

\thispagestyle{empty} \pagestyle{empty}
% numbers of pages are supplemented by the editor
%
% THE BEGINNING OF THE TEXT

\section{Introduction}
\vspace{-4pt}

The Penna model \cite{penna} is a popular tool for simulating aging in biological populations \cite{smdeoliveira98}. Since its introduction in 1995, more than 200 papers have been written utilizing the model \cite{pennabogota}. Much of its success is attributed to its simplicity and ability to reproduce  universal features of much more complicated, real phenomena \cite{stauffer&co} such as Gompertzian mortality \cite{smdeoliveira98,puhl&co}, catastrophic senescence of the Pacific salmon \cite{pennasalmon}, sexual reproduction \cite{smdeoliveira04}, speciation \cite{smdeoliveira&co1,luz-burgoa&co}, knowledge \cite{bustillos&oliveira}, and so forth. The model is based on the mutation accumulation theory of senescence which states that the strength of natural selection declines with age and a population can accumulate harmful mutations that only have late life effects \cite{smdeoliveira&co2}. That is, individuals die from bad genes which reveal or express themselves at later stages in life. 

The effects of finite resources are usually taken into account by setting a maximum sustainable population size - the environmental carrying capacity, $K$. It is generally believed that organisms compete for available food, space, and other necessities in order to stay alive. Thus, in computer simulations, the carrying capacity concept usually takes the form of a survival probability, also known as the Verhulst factor, $V=1-N(t)/K$. Here, $N(t)$ is the size of the population at time $t$. The Verhulst factor has been used to demonstrate different phenomena such as the Eve effect \cite{makowiec&co} and coevolution \cite{dabkowski&co}. But its main function is to limit growth \cite{raab}.

In the original Penna model \cite{penna,penna&stauffer}, the Verhulst term kills at random, regardless of fitness and age - VA implementation. Because random deaths in nature hardly play a significant role in population dynamics, this approach is not very well justified \cite{smdeoliveira&co3}. An alternative was later introduced applying the Verhulst factor to newborns only (VB implementation) \cite{niewczas&co}. This was done to avoid the accidental killing of healthy individuals \cite{martins&cebrat}. Furthermore, since majority of the population consists of newborns, limiting its size is enough to prevent exponential growth. 

Whether VA or VB, a greater part of the population is killed because of the Verhulst factor. The effectiveness of invoking a finite carrying capacity in capping simulated populations is undeniable. However, extra caution must be made so as not to confuse Verhulst effects with particular model features; otherwise, results can be misleading \cite{martins&cebrat}. 

This work investigates the consequences of varying Verhulst influence on the average characteristics of simulated populations. In particular, we examine the age demographics and survival rates of the steady state populations resulting from VA and VB implementations of the Penna model. 

\section{The model}
\vspace{-4pt}

The Penna model incorporates age structure in conventional population modeling via bit-handling techniques \cite{penna}. Here, individual characteristics are stored in a string of binary numbers (genome) that is defined upon birth. The string contains genetic information - zeroes are healthy genes and ones are bad genes or diseases. It is then read in sequence at a rate of one bit per time interval. This is done at the beginning of each iteration. Whenever a new bit is read, the individual's age is increased by one. A one on the $i$th bit means that the individual will experience the effects of a disease starting at age $i$.

An individual suffers a genetic death when the sum of active harmful mutations reaches the threshold value, $T$. The death age is determined for each individual at birth. When the total number of bad mutations in the genome is less than $T$, the individual dies at age equivalent to the bit-string length, $L$. Hence, the length of the string dictates the maximum lifespan. The concept of finite resources introduce random deaths into the simulation. Thus, for some, actual death may happen at an earlier age due to the Verhulst factor.

Surviving individuals reproduce only once, at age $R$. The number of offspring generated by each parent is given by the birth rate, $B$. The newborns copy the genes of the parent and acquire one additional mutation that is set at a random location. In the study of the influence of heredity and mutations on the evolution of a population, only those with deleterious effects are often times considered. This is because harmful mutations are many times more frequent in nature than beneficial ones \cite{pamilo&co}.

\begin{figure}
  \includegraphics[width=3in,angle=0]{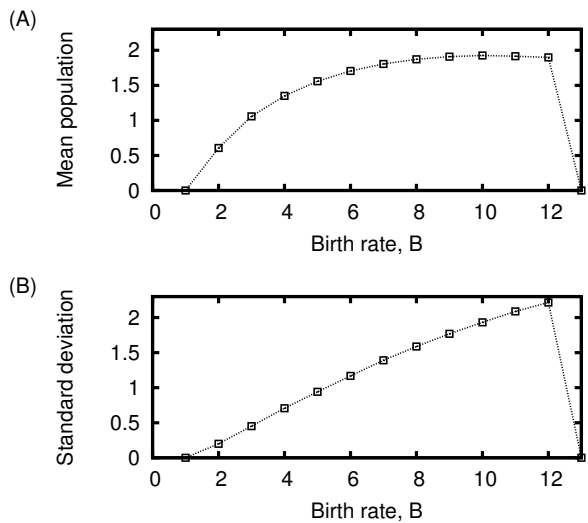}
\caption{(A) Mean population at steady state and (B) the corresponding standard deviations relative to the initial size ($N_0=20,000$) of populations obtained from a VA implementation with $T=3$ and $R=8$.}
\label{fig1}
\end{figure}

\begin{figure}
  \includegraphics[width=3in,angle=0]{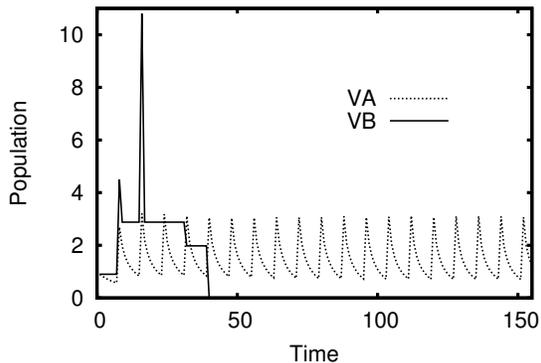}
\caption{Time evolution of the population associated with $B=4$, $T=3$ and $R=8$. Values are plotted as factors of the initial size, $N_0$. Note that in our simulations, $K=10 \times N_0$.}
\label{fig2}
\end{figure}

{\bf Simulation details:} Genetic information is stored in 32-bit long genomes ($L=32$). We begin our simulations with 20,000 perfect newborns (no bad genes). For VA and VB runs, the value of the carrying capacity is set ten times the initial population ($K=200,000$). At each iteration, age is updated first and deaths come immediately after. This procedure sets a maximum reproduction age, $R_{max}=L-1$, and for the VB case, restricts the action of the Verhulst factor to those at age 1. To minimize fluctuations, demographic data presented in the next section are averaged over the last 300 iteration. At these times, simulated populations are already at equilibrium.

\section{Results and Discussion}
\vspace{-4pt}

In Penna simulations, saturation time is influenced by the reproduction age. Those with higher $R$ values take longer to equilibriate \cite{penna}. The mutation threshold and the birth rate, on the other hand, affect population size. Total population generally increases with organism's tolerance for deleterious mutations. Figure~\ref{fig1}A shows a rise in mean population with birth rate. For these VA runs, the values saturate just under twice the initial population size. The magnitude of fluctuations (given by the standard deviation in Fig.~\ref{fig1}B) increases linearly with birth rate. The plots also show a maximum $B$ value that yields a nonzero steady state. Large fluctuations associated with high birth rates sometimes lead to extinction. A sudden rise in size, beyond the set carrying capacity, results in the accidental killing of the entire population. The same is observed in VB simulations. Fluctuations in the transient part cause premature extinction even at relatively lower birth rates (Fig.~\ref{fig2}). Note that although the effect of the Verhulst factor is restricted to a single age group, the strength of this external death term is not diminished. In fact, the number of random deaths is doubled in the VB case \cite{martins&cebrat}. Thus, the range of $B$ values that result in nonzero steady states is smaller for the VB implementation. It is for this reason that later comparisons of VA and VB simulation runs use different $B$ values.

\begin{figure}
  \includegraphics[width=3in,angle=0]{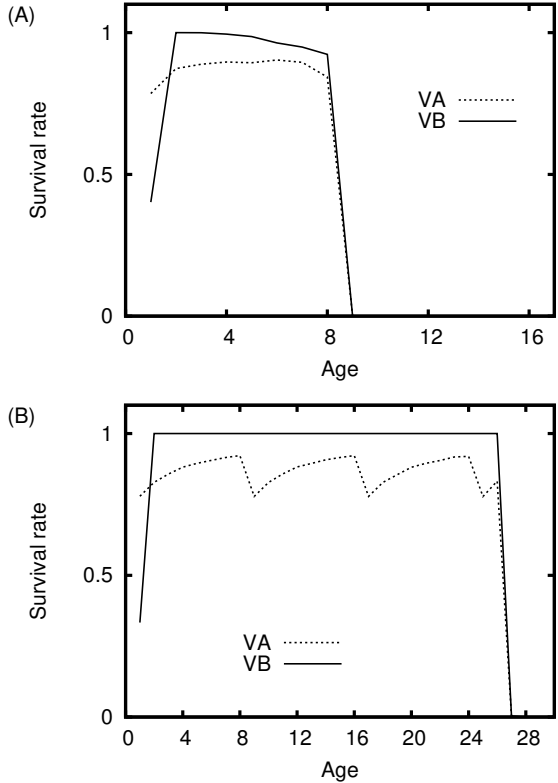}
\caption{Behavior of the survival rates when $B=3$, $R=8$: (A) $T=3$ and (B) $T=27$.}
\label{fig3}
\end{figure}

Recall from Section 2 that we restrict reproduction to a single age $R$. This one-time breeding strategy is characteristic of semelparous organisms. The aging process associated with semelparity is generally catastrophic \cite{pennasalmon,smdeoliveira04}. For the case of reproduction occuring only once at $R=8$, we plot the survival rates as a function of age (Fig.~\ref{fig3}). In aging studies, the survival rate\footnote{the ratio between the number of individuals with age $a$ at time $t$ and the number of individuals with age $a-1$ at time $t-1$, $N_a(t)/N_{a-1}(t-1)$} is related to organism fitness. The abrupt decline in fitness happens at age $R+1$, immediately after reproduction, when $T \le R$; otherwise, when $T>R$, the drastic loss in functional abilities is observed at the age equivalent to the threshold \cite{pinol&banzon2}. Notice further that the survival rates associated with the VA case in Fig.~\ref{fig3}B appear to have a period of $R$. This behavior is not a general feature of the model, but rather a consequence of our choice of initialization and Verhulst implementation \cite{pinol&banzon2}. The dips correspond to times when the Verhulst killing effect is strongest, felt by ages present at times when the population size is largest.

\begin{figure}
  \includegraphics[width=3in,angle=0]{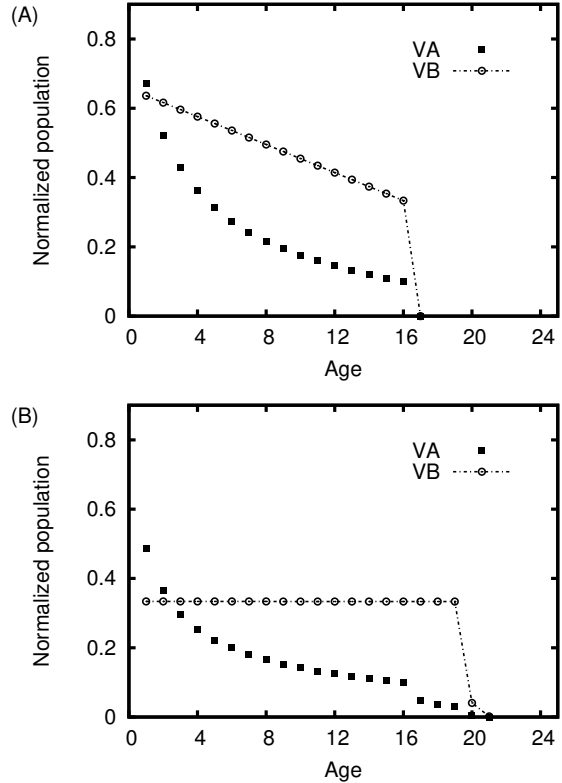}
\caption{Distribution of the population to its component ages, normalized about the average number of births. Here, $R=16$: (A) $T=1$ and (B) $T=20$. The $B$ values used are 10 and 3 for the VA and VB simulations, respectively.}
\label{fig4}
\end{figure}

The Penna model described in \cite{penna,penna&stauffer} follows Gompertz law \cite{smdeoliveira98,puhl&co} which predicts an exponential increase in mortality with age. Consequently, the distribution of the population to its component ages shows an exponential decrease. Note that these populations are VA and breed repeatedly, starting at age $R$ until death. For our current implementation, we obtain the same exponential decrease in the age demographics when the Verhulst factor is applied to all ages. However, when random deaths imposed by the Verhulst factor is restricted to newborns only, we lose the Gompertzian structure (Fig.~\ref{fig4}). 

\begin{figure}
  \includegraphics[width=3in,angle=0]{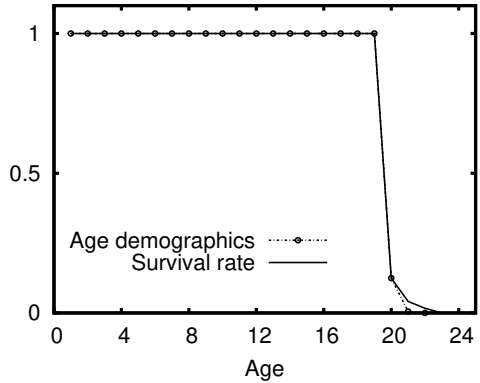}
\caption{Demographic statistics obtained for the Verhulst-free case: $B=1, T=20$ and $R=16$.}
\label{fig5}
\end{figure}

Figure~\ref{fig5} presents the demographic data associated with Verhulst-free populations. These are results of simulation runs that do not account for environmental restrictions. That is, there are no random killings by the Verhulst factor and all deaths are due to genetic reasons. Age demographics and survival rates obtained for the Verhulst-free case show no vartiation in mortality rate and physiological condition with age, respectively. These are two of the criteria identified by Caleb Finch \cite{finch} for negligible senescence. These describe the very slow aging process observed in coldwater fish, bivalves, turtles, whales, and, fairly recently, in the naked mole-rats \cite{buffenstein}. Note that within the framework of the Penna model, Verhulst-free demonstrations are limited to threshold values that are greater than the reproduction age ($T>R$) with one semelparous birth per adult ($B=1$) \cite{pinol&banzon1}. So far, these are the only parameter sets that lead to nonzero steady states even without random deaths by the Verhulst factor.

\section{Summary and conclusion}
\vspace{-4pt}

Population dynamics models provide quantitative estimates of the population size at some time $t$, given its initial size, growth rate, and the carrying capacity \cite{martinez&co}. The carrying capacity incorporates all possible interactions among individuals in an environment with limited resources into a single parameter. Its value determines the size of the steady state population \cite{murray}. Without the concept of a carrying capacity, populations behave exponentially. Thus, most models deem this parameter necessary to avoid the exponential growth.

In computer models, the concept generally takes the form of a Verhulst factor - a random death probability. Despite the fact that majority of deaths in simulations is attributed to this term, its effect on the properties of simulated populations is rarely considered. The choice of implementation of this random death strategy unfortunately has an unexpected impact on the population's genetic profile \cite{martins&cebrat}. In addition to this, we find that restricting the action of the Verhulst term to newborns only (VB implementations) alters some of the basic model features which was demonstrated imposing the death probability to all individuals regardless of fitness and age (VA implementation).

\vspace{10pt} \noindent
{\bf Acknowledgement:} \ The research was supported in part by the Department of Science and Technology (DOST) through the Accelerated Science and Technology Human Resource Development Program (ASTHRDP).


\begin{thebibliography}{11}

\vspace{-7pt}
\bibitem{penna}
T.J.P.~Penna, A bit-string model for biological aging, {\em J.~Stat.~Phys.} 78, 1995, pp.~1629--1633.

\vspace{-7pt}
\bibitem{smdeoliveira98}
S.~Moss~de~Oliveira, A small review of the Penna model for biological ageing, {\em Physica~A} 257, 1998, pp.~465--469.

\vspace{-7pt}
\bibitem{pennabogota}
T.J.P.~Penna, Aging and evolution, http://profs.if.uff.br/tjp/palestras/bogota.pdf, Accessed 23 Nov 2010.

\vspace{-7pt}
\bibitem{stauffer&co}
D.~Stauffer, P.M.C.~de~Oliveira, S.~Moss~de~Oliveira, T.J.P.~Penna and J.S.~S$\acute{a}$~Martins, Computer simulations for biological ageing and sexual reproduction, {\em An.~Acad.~Bras.~Cienc.} 73, 2001, pp.~15--32.

\vspace{-7pt}
\bibitem{puhl&co}
H.~Puhl, D.~Stauffer and S.~Roux, Ageing, war and predators, {\em Physica~A} 221, 1995, pp.~445--452.

\vspace{-7pt}
\bibitem{pennasalmon}
T.J.P.~Penna, S.~Moss~de~Oliveira and D.~Stauffer, Mutation accumulation and the catastrophic senescence of the Pacific salmon, {\em Phys.~Rev.~E} 52, 1995, pp.~R3309--R3312.

\vspace{-7pt}
\bibitem{smdeoliveira04}
S.~Moss~de~Oliveira, Evolution, ageing and speciation: Monte Carlo simulations of biological systems, {\em Braz. J. Phys.} 34, 2004, pp.~1066--1076.

\vspace{-7pt}
\bibitem{smdeoliveira&co1}
S.~Moss~de~Oliveira, J.S.~S$\acute{a}$~Martins, P.M.C.~de~Oliveira, K.~Luz-Burgoa, A.~Ticona and T.J.P.~Penna, The Penna model for biological aging and speciation, {\em Computing~in~Science~\&~Engineering} 6, 2004, pp.~74--81.

\vspace{-7pt}
\bibitem{luz-burgoa&co}
K.~Luz-Burgoa, S.~Moss~de~Oliveira, J.S.~S$\acute{a}$~Martins, D.~Stauffer and A.O.~Sousa, Computer simulation of sympatric speciation with Penna ageing model, {\em Braz. J. Phys.} 33, 2003, pp.~623--627.

\vspace{-7pt}
\bibitem{bustillos&oliveira}
A.~Ticona~Bustillos and P.M.C.~de~Oliveira, Evolutionary model with genetics, aging and knowledge, {\em Phys. Rev. E} 69, 2004, pp.~021903.1--021903.9.

\vspace{-7pt}
\bibitem{smdeoliveira&co2}
S.~Moss~de~Oliveira, D.~Alves and J.S.~S$\acute{a}$~Martins, Evolution and ageing, {\em Physica~A} 285, 2000, pp.~77--100.

\vspace{-7pt}
\bibitem{makowiec&co}
D.~Makowiec, J.~D\c{a}bkowski and M.~Groth, The Eve effect in the Penna model of biological ageing, {\em Physica~A} 273, 1999, pp.~169--181.

\vspace{-7pt}
\bibitem{dabkowski&co}
J.~D\c{a}bkowski, M.~Groth and D.~Makowiec, Verhulst factor in the Penna model of biological aging, {\em Acta~Phys~Polonica~B} 31, 2000, pp.~1027--1035.

\vspace{-7pt}
\bibitem{raab}
A.~Raab, Comment on \lq\lq A bit-string model for biological aging\rq\rq, {\em J.~Stat.~Phys.} 91, 1998, pp.~1055--1060.

\vspace{-7pt}
\bibitem{penna&stauffer}
T.J.P.~Penna and D.~Stauffer, Efficient Monte Carlo simulation of biological aging, {\em Int.~J.~Mod.~Phys.~C} 6, 1995, pp.~233--239.

\vspace{-7pt}
\bibitem{smdeoliveira&co3}
S.~Moss~de~Oliveira, P.M.C.~de~Oliveira and J.S.~S$\acute{a}$~Martins, Penna bit-string model with constant population, {\em Int.~J.~Mod.~Phys.~C} 15, 2000, pp.~301--305.

\vspace{-7pt}
\bibitem{niewczas&co}
E.~Niewczas, S.~Cebrat and D.~Stauffer, The influence of medical care on the human life expectancy in $20^{th}$ century and the Penna ageing model, {\em Theory~Biosci.} 119, 2000, pp.~122--131.

\vspace{-7pt}
\bibitem{martins&cebrat}
J.S.~S$\acute{a}$~Martins and S.~Cebrat, Random deaths in a computational model for age-structured populations, {\em Theory~Biosci.} 119, 2000, pp.~156--165.

\vspace{-7pt}
\bibitem{pamilo&co}
P.~Pamilo, M.~Nei and W.-H.~Li, Accumulation of mutations in sexual and asexual populations, {\em Genet.~Res.,~Camb.} 49, 1987, pp.~135--146.

\vspace{-7pt}
\bibitem{pinol&banzon2}
C.M.~Pi\~nol and R.~Banzon, Catastrophic senescence and semelparity in the Penna aging model, {\em Theory Biosci.} DOI 10.1007/s12064-010-0115-7.

\vspace{-7pt}
\bibitem{finch}
C.~Finch, {\em Longevity, senescence and the genome,} University of Chicago Press, Chicago 1990 as cited in \cite{buffenstein}.

\vspace{-7pt}
\bibitem{buffenstein}
R.~Buffenstein, Negligible senescence in the longest living rodent, the naked mole-rat: insights from a successfully aging species, {\em J. Comp. Physiol. B} 178, 2008, pp.~439--445.

\vspace{-7pt}
\bibitem{pinol&banzon1}
C.M.~Pi\~nol and R.~Banzon, Stability in a population model without random deaths by the Verhulst factor, to be published in {\em Physica A}.

\vspace{-7pt}
\bibitem{martinez&co}
A.S.~Martinez, B.C.T.~Cabella and F.~Ribeiro, Scaling function, universality and analytic solutions of generalized one-species population dynamics models, e-print arXiv:1010.2950v1.

\vspace{-7pt}
\bibitem{murray}
J.D.~Murray, {\em Mathematical Biology: I. An introduction (3rd ed),} Springer --Verlag, Berlin --Heidelberg 2002.

\end{thebibliography}
\end{document}